\documentclass[12pt]{article}
\usepackage[a4paper, total={7in, 10in}]{geometry}
\usepackage[parfill]{parskip}
\usepackage{physics, tensor, float, subcaption}
\usepackage{graphicx}
\graphicspath{ {Plots/} }
\usepackage{jhep-mod}
\usepackage{bm}
\usepackage{soul}
\usepackage{amssymb,amsmath,amsthm}
\usepackage{mathrsfs}
\usepackage[utf8]{inputenc}
\usepackage{enumerate}
\usepackage{bigints}
\usepackage{xcolor}
\usepackage{appendix}
\usepackage{graphicx}
\usepackage{float}
\usepackage{tikz}
\usepackage{setspace}
\usepackage{cancel}
\usepackage{doi}
\definecolor{purple}{rgb}{1,0,1}
\definecolor{lime}{HTML}{A6CE39} 

\definecolor{lime}{HTML}{A6CE39}
\newcommand{\orcidicon}{%
	\begin{tikzpicture}
	\draw[lime, fill=lime] (0,0) 
		circle [radius=0.16] 
		node[white] {{\fontfamily{qag}\selectfont \tiny ID}};
	\draw[white, fill=white] (-0.0625,0.095) 
		circle [radius=0.007];
	\end{tikzpicture}
	\hspace{-5mm}
}
\newcommand\orcidMatt{{\href{https://orcid.org/0000-0003-1088-6485}{\orcidicon}}}

\renewcommand{\O}{\mathcal{O}}
\begin{document}

\title{\huge{
Efficient computation of null affine parameters}}

\author{
\Large
Matt Visser\!\orcidMatt\!
}
\affiliation{School of Mathematics and Statistics, Victoria University of Wellington, \\
\null\qquad PO Box 600, Wellington 6140, New Zealand.}
\emailAdd{matt.visser@sms.vuw.ac.nz}
\def\theta{\vartheta}
\def\O{{\mathcal{O}}}
\def\d{{\mathrm{d}}}

\abstract{
\vspace{1em}

Finding affine parameters for null geodesics is often of considerable physical importance, especially when studying the paths swept out by null geodesics or dealing with conservation laws and/or averaged energy conditions. But explicitly finding null affine parameters is also often quite tedious and can sometimes even be somewhat tricky.
Herein we shall demonstrate that the existence of a conformally related spacetime containing a conformal Killing vector, timelike in the domain of outer communication, is quite sufficient to define a preferred set of spatial three-slices---on which a well-defined ``affine'' three-metric can be introduced to capture the notion of affine null parameter---\emph{before} 
  explicitly finding the null geodesics. The construction depends on the properties of conformal transformations and on the conserved quantity associated with the  conformal Killing vector.  Having the affine null parameter in hand before attempting to find the actual null geodesics often quite radically simplifies other parts of the analysis. 
We emphasize that the successful identification of affine null parameters is a general-purpose tool of wide applicability in both general relativistic and astrophysical settings.

\bigskip

\bigskip
\noindent
{\sc Date:} 15 November 2022; 20 December 2023;  \LaTeX-ed \today

\bigskip
\noindent{\sc Keywords}: \\
null geodesics, affine parameters, Killing vectors, conformal Killing vectors; \\
conformal transformations. 

\bigskip
\noindent{\sc arXiv}:   2211.07835 [gr-qc]

\bigskip
\noindent{\sc Published as}:  Universe {\bf9} (2023) 521.
}

\maketitle
\def\d{{\mathrm{d}}}
\def\L{{\mathcal{L}}}
\parindent0pt
\parskip7pt



\bigskip
\clearpage

\hrule\hrule\hrule

\bigskip

\section{Introduction}

Null curves, and more specifically null geodesics, and the affine parameters that are always associated with null geodesics, are standard tools in both general relativity and astrophysics~\cite{MTW, Wald,Carroll,Hartle,Schutz1,Weinberg1,Weinberg2, Peebles, Padmanabhan, Hobson,Sachs,Stephani, Dinverno,Poisson,Lorentzian}. At a technical level:
For any smooth null curve $x^a(\lambda)$ with a null tangent vector $k^a= \d x^a/\d\lambda$ and some arbitrary parameterization $\lambda$, the {general} geodesic equation is $\nabla_k k \propto k$.  It is only for the so-called affine parameterizations that the geodesic equation {specifically} reduces to $\nabla_k k =0$. See, for instance, any of the references~\cite{MTW, Wald,Carroll,Hartle,Schutz1,Weinberg1,Weinberg2, Peebles, Padmanabhan, Hobson,Sachs,Stephani,Dinverno,Poisson,Lorentzian}.\footnote{{The situation for nongeodesic null curves is, if anything, worse. Changing the parameterization of a nongeodesic null curve will again modify the ``acceleration'' $\nabla_k k$ by terms proportional to the tangent vector $k$, but there is now no longer a simple prescription for choosing a preferred affine parameterization.}}
 In any {fixed but} arbitrary spacetime, once a specific null geodesic has been found, finding the corresponding affine parameter is a straightforward if tedious textbook exercise---amounting to solving a first-order ordinary differential equation~\cite{MTW, Wald,Carroll, Hartle,Schutz1,Weinberg1,Weinberg2, Peebles, Padmanabhan, Hobson,Sachs, Stephani, Dinverno,Poisson,Lorentzian}. But can this process be simplified and rearranged? Under what conditions is it possible to choose an affine parameter first, \emph{{before}} explicitly finding the null geodesics?

Identifying an appropriate affine null parameter is essential, for instance, when one is attempting to implement the averaged null energy condition (ANEC) or any of its variants~\cite{Flanagan:1996, Yurtsever:1990, Wald:1991, Borde:1987, Ford:1994, Ford:1995, Graham:2007, Wall:2009wi, Kontou:2015yha, Capozziello:2013, Capozziello:2014, Curiel:2014, Hochberg:1998, Barcelo:2002, Fewster:2002, Kar:2004, Martin-Moruno:2017, Visser:1994, Visser:1999, Yurtsever:1995, Yurtsever:1994wc, Fewster:2006, Graham:2005, Kontou:2015,Penrose:1993ud,Morris:1988tu, Witten:2019}. 
{The crucial point is that for the arbitrary parameterizations of the null curve, the ANEC is essentially content-free since the ANEC integral could then be arbitrarily adjusted by simply changing the geodesic parameterization. Only for affine parameters, since the relation between differing affine parameters is then very tightly constrained  {(via an affine rescaling $\lambda\to a\lambda +b$)}, is the positivity of the ANEC an invariant concept.
Carefully formulating the ANEC (or its variants)} is, in turn, essential for understanding the relevant singularity theorems~\cite{Witten:2019, Senovilla:1998, Hell, Roman:1983, Roman:1986tp, Roman:1988, Bojowald:2007, Parker:1973, Fewster:2010, Abreu:2010, Ford:2003}, 
topological censorship~\cite{Friedman:1993,Galloway:1999,Jacobson:1994, Shapiro:1995, Chrusciel:2009, Browdy:1995, Eichmair:2012, Chrusciel:2019},
and/or Hawking's chronology protection conjecture~\cite{Hawking:1991, Visser:1992comments, Kay:1996, Visser:1996ring, Visser:1997, Visser:2002hawking60, Friedman:2006, Liberati:2016mess}. 
Further afield, the use of null affine parameters is extremely common in ongoing research, both theoretical and observational, such as, for instance, in~\cite{Eling:2006df, Olmo:2015axa, Hollands:2019whz,Bejarano:2017fgz, Ashtekar:2010qz, Menchon:2017qed, Flanagan:2018yzh, Maeda:2010ja, Yang:2013yoa, Sanghai:2017yyn, Arrechea:2021pvg, Fuentes:2019nel, Anastopoulos:2020mrt, White:2022paq, Boyanov:2023uzi, Arrechea:2023hmo, Dexter:2009fg, Jacobson:2007tj, Chandrasekaran:2019ewn, Chan:2017igo, Preston:2006zd}.

Herein we shall demonstrate that the existence of a conformally related spacetime containing a (conformal) Killing vector, timelike in the domain of outer communication, is quite sufficient to allow us to make some quite general and powerful observations in terms of an ``affine'' three-metric defined on the spatial slices. 

We shall consider both the usual ADM slicings, see~\cite{ADM,Gourgoulhon:2007, Hawking:1995} and {various textbook presentations such as}~\cite{MTW,Wald,Carroll,Hartle,Schutz1,Weinberg1,Weinberg2, Peebles, Padmanabhan, 
Hobson,Sachs,Stephani,Dinverno,Poisson,Lorentzian}, and also the complementary ``threaded''  (Kaluza--Klein-like) formulations, see~\cite{thread, vanElst:1996,Ellis:1998}.

After some preliminary comments on generalities and nonaffine parameterizations, let us look at six cases of increasing levels of generality:
\begin{itemize}
\item Manifestly ultrastatic spacetimes.
\item Manifestly static spacetimes.
\item Stationary spacetimes.
\item Spacetimes with a timelike conformal Killing vector.
\item Conformal deformations.
\item The completely general case.
\end{itemize}

As long as there is at least some symmetry, one can be reasonably explicit regarding the affine parameter. 
But ultimately, we shall see that in the completely general case, one can at best nonconstructively argue for the \emph{existence} of a purely formal {Finslerian}
\mbox{\cite{Bekenstein:1992, Perlick:2005, Gibbons:2007,Gibbons:2008,Skakala:2010, Lammerzahl:2012, Lammerzahl:2018, Pfeifer:2019} result.}


\section{Generalities}
In complete generality, for any {differentiable} vector field $K^a(x)$ and any curve with tangent vector $k^a(\lambda)$, we have the standard mathematical identity
\begin{equation}
{\d\over\d\lambda} \left[g\left(K, {\d x^a \over \d \lambda } \right)\right] =  {\d\over\d\lambda}\left[ K_a k^a\right] = 
K_{a;b} {\d x^b\over\d\lambda} k^b + K_a {\d\over\d\lambda}k^a = K_{(a;b)} \;k^a k^b + K_a k^a{}_{;b} k^b.
\end{equation}
l{That is,} 
\begin{equation}
\label{E:null0}
{\d\over\d\lambda} \left[g\left(K, {\d x^a \over \d \lambda } \right)\right] 
= K_{(a;b)} \;k^a k^b + K_a (\nabla_k k)^a.
\end{equation}
{We} 
 shall systematically apply and reapply this general identity in the discussion below. 
In particular, for an affinely parameterized geodesic (either null or timelike), since $\nabla_k k=0$, we have~\cite{MTW,Wald,Carroll,Hartle, Schutz1, Weinberg1,Weinberg2, Peebles, Padmanabhan, Hobson,Sachs,Stephani,Dinverno,Poisson,Lorentzian}
\begin{equation}
\label{E:generic}
{\d\over\d\lambda} \left[g\left(K, {\d x^a \over \d \lambda } \right)\right] 
= K_{(a;b)} \; k^a k^b.
\end{equation}
{Ultimately,} 
 it is the presence of the quantity $K_{(a;b)}$ in this {purely mathematical} identity that focuses attention on Killing vectors, on conformal Killing vectors, and ultimately on spacetimes conformal to spacetimes containing a conformal Killing vector.

\section{Nonaffine Parameterizations}
It is sometimes {(albeit somewhat rarely)} useful to adopt a nonaffine parameterization. Let us see what happens in this situation, and see what we are seeking to \emph{avoid} with the approach to be developed below. For a null geodesic with the tangent vector $k^a(\lambda)=\d x^a(\lambda)/\d\lambda$ that is not affinely parameterized, one has $\nabla_k k = f(\lambda) \, k(\lambda)$, so that the general identity (\ref{E:null0}) becomes
\begin{equation}
{\d\over\d\lambda} \left[g\left(K, {\d x^a \over \d \lambda } \right)\right] 
= K_{(a;b)} \;k^a k^b + f(\lambda) \; K_a k^a.
\end{equation}
{This} 
 can be rewritten as 
\begin{equation}
\label{E:generic2}
 \exp\left(+ \int^\lambda f(\bar\lambda) \d\bar \lambda \right)
{\d\over\d\lambda} \left[ \exp\left(- \int^\lambda f(\bar\lambda)\, \d\bar \lambda \right) \; g\left(K, {\d x^a \over \d \lambda } \right)\right] 
= K_{(a;b)} \; k^a k^b.
\end{equation}
{Comparing} 
 this with the affine case of Equation (\ref{E:generic}), we see that (up to an irrelevant  overall constant normalization), we can set
\begin{equation}
{d\lambda\over\d\lambda_\mathrm{affine}} =  \exp\left(- \int^\lambda f(\bar\lambda)\, \d\bar \lambda \right),
\end{equation}
whence
\begin{equation}
\lambda_\mathrm{affine}(\lambda) = 
\int^\lambda \exp\left(+ \int^{\tilde\lambda} f(\bar\lambda)\,\d\bar \lambda \right)\; \d\tilde\lambda.
\end{equation}
{While} 
 this can always be constructed, it is precisely the need for evaluating this somewhat messy double-iterated integral that we will seek to evade developing the formalism and discussion below.

Given the above discussion, one might reasonably ask whether or not nonaffine parameterizations are ever useful? 
One specific situation in which nonaffine parameters are arguably useful is in working with coordinates specifically adapted to match measurements carried out in a laboratory setting. 
To see how this works, recall that the equation for a geodesic in arbitrary nonaffine parameterization is
\begin{equation}
\label{E:geodesic}
{\d^2 x^a\over \d\lambda^2} + \Gamma^a{}_{bc} \; {\d x^b\over \d\lambda} \; {\d x^c\over \d\lambda}
= f(\lambda) \; {\d x^a\over \d\lambda}.
\end{equation}
{Now} 
 suppose that we can, in some region, ``naturally'' assign a
time-coordinate $t$ and three spatial coordinates $x^i$ and adopt a coordinate patch (chart) of the form (for convenience, we set $c=1$):
\begin{equation}
x^a = (t, \;x^i).
\end{equation}
{This} 
is an extremely mild condition: we are simply choosing a coordinate patch in which one of the coordinates is timelike---and this can \emph{{always}} be conducted locally. For instance, this is exactly what one would do in a laboratory setting.
But this condition is more general than merely applying to laboratory situations and might also be profitably adopted for investigating solar system physics. At least locally, it is truly a generic choice, not a physical constraint. 

Let us now agree to parameterize all curves by the nonaffine parameter $t$, denote coordinate derivatives with respect to $t$ by an overdot,\footnote{{We emphasize: The overdots are not proper time derivatives or affine derivatives. The overdots are intimately entangled with the specific choice of coordinate system.}}$^,$\footnote{{We emphasize: this is a choice, not a physical constraint.}} and so define
\begin{equation}
\label{E:geodesic}
{\d x^a\over \d t} = 
\left( {\d t\over \d t} ; {\d x^i\over\d t} \right) = 
\left( 1; {\dot x^i} \right).
\end{equation}
Then, if we split the geodesic Equation (\ref{E:geodesic}) into time+space components, we find, from the first (zeroth) component, that
\begin{equation}
{\d^2 t\over \d t^2} + \Gamma^t{}_{bc} \; {\d x^b\over \d t} \; {\d x^c\over \d t} 
= f(t) \; {\d t\over \d t},
\end{equation}
whence
\begin{equation}
0  
+ \Gamma^t{}_{tt} 
+ \left[ \Gamma^t{}_{ti} +\Gamma^t{}_{it} \right] \dot x^i 
+ \Gamma^t{}_{ij} \; \dot x^i\;\dot x^j
= f(t).
\end{equation}
That is, in these specific coordinates and with this specific choice of parameterization, the function $f(t)$ is completely specified by the equation
\begin{equation}
f(t) = \Gamma^t{}_{bc} \; {\d x^b\over \d t} \; {\d x^c\over \d t}. 
\end{equation}
Equivalently,
\begin{equation}
f(t) = 
\Gamma^t{}_{tt} + \left[ \Gamma^t{}_{ti} +\Gamma^t{}_{it} \right] \dot x^i+ \Gamma^t{}_{ij} \; \dot x^i\;\dot x^j.
\end{equation}
So $f(t)$ is, in this framework, quadratic in the coordinate three-velocity.

Now substituting this back into the spatial part of the geodesic Equation (\ref{E:geodesic}), we see
\begin{equation}
{\d^2 x^i\over \d t^2} + \Gamma^i{}_{bc} \; {\d x^b\over \d t} \; {\d x^c\over \d t} 
= f(t) \; {\d x^i\over \d t}.
\end{equation}
thence, the coordinate three-acceleration is 
\begin{equation}
\ddot x^i
= - \Gamma^i{}_{tt} - 
\left[ \Gamma^i{}_{tj} + \Gamma^i{}_{jt} \right]  \dot x^j -  \Gamma^i{}_{jk} \; \dot x ^j \; \dot x^k 
+ \left[  \Gamma^t{}_{tt} + \left[ \Gamma^t{}_{tj} +\Gamma^t{}_{jt} \right] \dot x^j + \Gamma^t{}_{jk} \; \dot x^j\;\dot x^k\right] \dot x^i. 
\end{equation}
This can be rewritten as
\begin{equation}
\ddot x^i
= - \Gamma^i{}_{tt} - 
\left[ \Gamma^i{}_{tj} + \Gamma^i{}_{jt} 
-\Gamma^t{}_{tt}\; \delta^i{}_j \right]  \dot x^j 
-  \left[\Gamma^i{}_{jk} 
-  (\Gamma^t{}_{tj} +\Gamma^t{}_{jt} ) \delta^i{}_k \right]  \dot x^j \; \dot x^k 
+ \left[\Gamma^t{}_{jk} \right]\; \dot x^j\;\dot x^k\; \dot x^i.
\end{equation}
Thus, in this coordinate system, with this choice of parameterization,  a geodesic experiences a (cubic) 
velocity-dependent (but nevertheless composition-independent and universal)
``coordinate 3-acceleration'' . This is \emph{{not}} the three-acceleration; it is a coordinate three-acceleration. 
We are looking at a geodesic, so the four-acceleration is identically zero.\footnote{{The only even mildly surprising part of the result is the presence in the three-acceleration of a term cubic in the three-velocity.}}

If the object in question is a massive particle that is momentarily at rest ($\dot x^i=0)$, then
\begin{equation}
\ddot x^i
\to  - \Gamma^i{}_{tt}.
\end{equation}
Note that this statement is \emph{{exact}}; it relies on picking an
appropriate coordinate system in which the particle is momentarily at
rest but does not make any weak-field approximation.
As a counterpoint, for a massless particle such as a photon, we always have $\dot x^i \neq 0$, and the three-velocity dependent terms cannot be neglected. 
This gives a \emph{{qualitative}} and \emph{{generic}} explanation for why the three-acceleration of photons differs from the three-acceleration of planets.
With a little extra work, the discussion can be made fully \emph{{quantitative}} and \emph{{explicit}}. 

Overall, while nonaffine parameterizations are sometimes useful for specific purposes, they tend to be highly coordinate-dependent and are not particularly easy to generalize. For instance, in the current framework, we have the rather messy (but fully explicit) result 
\begin{equation}
\lambda_\mathrm{affine}(t) = 
\int^t \exp\left(+ \int^{\tilde t} \left\{
\Gamma^t{}_{tt} + \left[ \Gamma^t{}_{ti} +\Gamma^t{}_{it} \right] \dot x^i+ \Gamma^t{}_{ij} \; \dot x^i\;\dot x^j
\right\}_{\bar t} \;\;\d\bar t \right)\; \d\tilde t.
\end{equation}
In subsequent sections below, we shall use symmetries and partial symmetries to (as much as possible) side step the need for explicit computations of this type.

\section{Manifestly ultra-static spacetimes}

An ultrastatic spacetime is characterized by the global existence of a hypersurface-orthogonal timelike unit-norm Killing vector, satisfying $g(K,K)=-1$. See, for instance, references~\cite{Page:1982,Blau:1988,Zelnikov:1997,Furlani:1997fh,Popov:2004yk,Fewster:2003ey, Sonego:2010,Fewster:2013}. 
In any ultrastatic spacetime, one can (without a loss of generality) choose coordinates to make the metric block diagonal and the ultrastatic condition manifest, such that
\begin{equation}
g_{ab} = \left[\begin{array}{r|c}
- 1 & 0 \\ \hline 0 & g_{ij}
\end{array} \right].
\end{equation}
Here, the three-metric $g_{ij}(\vec x)$ is time-independent.
{We emphasize that while the metric can be put in this form, there is an infinity of other coordinate systems that do not share this property.  If the metric is presented in the form above, then we say that the ultrastatic symmetry is manifest. 
Working in other coordinate systems is possible but may or may not be particularly useful depending on the questions being addressed. In the manifestly ultrastatic coordinate system}
for any null curve, {we have}
\begin{equation}
-\d t^2 + g_{ij} \,\d x^i \, \d x^j = 0,
\end{equation}
implying (up to an unimportant sign)
\begin{equation}
\d t =  \sqrt{ g_{ij} \,\d x^i \, \d x^j }.
\end{equation}
Thence, in the current context, for any affinely parameterized null geodesic, one  has the
conservation law~\cite{MTW,Wald,Carroll,Hartle,Schutz1,Weinberg1,Weinberg2, Peebles, Padmanabhan, Hobson,Sachs,Stephani,Dinverno,Poisson,Lorentzian}
\begin{equation}
g\left(K, {\d x^a \over \d \lambda } \right) = 
g_{ta} {\d x^a \over \d \lambda } 
= - {\d t \over \d \lambda } = \hbox{constant} = \epsilon.
\end{equation}
Thence, along any affinely parameterized null geodesic in an ultrastatic spacetime, up to an undetermined constant, one has
\begin{equation}
\d \lambda \propto \d t =  \sqrt{  g_{ij} \,\d x^i\, \d x^j}.
\end{equation}
That is,
in any ultrastatic spacetime, the preferred time coordinate, or equivalently the spatial distance traveled, can be used as an affine parameter. 
While this particular ultrastatic situation is almost trivial, the discussion now readily generalizes to both static and stationary spacetimes---and beyond. 

\paragraph{Example:}

A specific class of examples of ultrastatic spacetimes that arises from outside the general relativity community is the class of Lorentzian metrics describing index-gradient methods in optics. Let $n(\vec x)$ be a position-dependent refractive index (achieved, for instance, by doping the background optical medium with various admixtures of other materials). 
Then, define the ultrastatic metric
\begin{equation}
g_{ij}(\vec x) = n^2(\vec x)\; \delta_{ij}; \qquad g_{ab}(\vec x) = -1 \oplus g_{ij}(\vec x).
\end{equation}
This ultrastatic metric adequately describes the propagation of light in the ray optics limit, and from the above discussion, we see
\begin{equation}
\d \lambda \propto \d t =  n(\vec x) \; || \d{}\vec x||.
\end{equation}
Thence, in index-gradient optics, the affine parameter is simply the optical distance.

\section{Manifestly static spacetimes}

A static spacetime is fully characterized by the global existence of a hypersurface orthogonal Killing vector $K$, which is timelike (that is, $g(K,K)<0$) in the domain of outer communication. See, for instance, references~\cite{MTW,Wald, Carroll, Hartle, Schutz1,Weinberg1,Weinberg2, Peebles, Padmanabhan, Hobson,Sachs,Stephani,Dinverno, Poisson, Lorentzian,Padmanabhan:2003, Perlick:2003,Visser:1992dbh, Hochberg:1997throat, Rahman:2001,Martin:2003}.
In a static spacetime, one can (without a loss of generality) choose coordinates to make the metric block diagonal and the static condition manifest, such that
\begin{equation}
g_{ab} = \left[\begin{array}{c|c}
- N^2 & 0 \\ \hline 0 & g_{ij}
\end{array} \right],
\end{equation}
where both $N(\vec x)$ and $g_{ij}(\vec x)$ are time-independent
(again, while this can be conducted, it is not strictly speaking necessary to do so. 
But making the symmetry manifest by the judicious choice of coordinates simplifies the discussion).
\enlargethispage{20pt}

Then, for any null curve
\begin{equation}
- N^2 \d t^2 + g_{ij} \d x^i \d x^j = 0,
\end{equation}
implying (up to an unimportant sign)
\begin{equation}
N \;\d t = \sqrt{g_{ij} \;\d x^i \d x^j}.
\end{equation}
For any affinely parameterized null geodesic, one 
has the conservation law
\begin{equation}
g\left(K, {\d x^a \over \d \lambda } \right) 
= g_{ta} {\d x^a \over \d \lambda } 
= - N^2 {\d t \over \d \lambda } = \hbox{constant} = \epsilon.
\end{equation}
Thence, along any affinely parameterized null geodesic, up to some undetermined arbitrary nonzero constant,  one has
\begin{equation}
\d \lambda \propto N^2 \d t =  N \sqrt{  g_{ij} \,\d x^i \,\d x^j} =
\sqrt{ N^2  g_{ij} \,\d x^i \, \d x^j}.
\end{equation}
This can also be written as
\begin{equation}
\d \lambda \propto - g_{tt}\, \d t = 
\sqrt{ - g_{tt}\, g_{ij} \,\d x^i \, \d x^j} = 
\sqrt{ h_{ij} \,\d x^i \, \d x^j}.
\end{equation}
Note carefully that the ``affine'' three-metric $h_{ij}= - g_{tt}\, g_{ij} $ appearing above is \emph{not} what is sometimes called the ``optical metric''. 
The optical metric would instead be the conformally related quantity 
\begin{equation}
g^{optical}_{ij} = {g_{ij}\over N^2} = - {g_{ij}\over g_{tt}}= {h_{ij}\over  g_{tt}^2}.
\end{equation}
While the optical metric is useful for determining paths of light rays, it is the affine three-metric that is useful for determining affine parameters. More on these issues later.


\paragraph{Examples:}


\begin{itemize}

\item Schwarzschild spacetime in curvature coordinates:\\ One has
\begin{equation}
\d s^2 = -(1-2m/r) \d t^2 + {\d r^2\over 1-2m/r} + r^2\d\Omega^2; 
\end{equation}
then
\begin{equation}
\d \lambda \propto \sqrt{ \d r^2 + (r^2-2mr) \d\Omega^2}.
\end{equation}
So in particular, for radial null geodesics, without any further calculation,  $\lambda \propto r$\\
(this is tolerably well-known, though it is usually checked by a brute-force analysis of the radial null geodesics).

\item Reissner--Nordstr\"om spacetime in curvature coordinates:\\ One has
\begin{equation}
\d s^2 = -(1-2m/r+q^2/r^2) \d t^2 
+ {\d r^2\over 1-2m/r+q^2/r^2} + r^2\d\Omega^2; 
\end{equation}
then
\begin{equation}
\d \lambda \propto \sqrt{ \d r^2 + (r^2-2mr+q^2) \d\Omega^2}.
\end{equation}
So in particular, for radial null geodesics, without any further calculation,  $\lambda \propto r$\\
(this is tolerably well-known, though it is usually checked by a brute-force  analysis of the radial null geodesics).

\item Static spherical symmetry (in curvature coordinates): \\
One has
\begin{equation}
\d s^2 = -e^{-2\Phi(r)} [1-2m(r)/r] \, \d t^2 
+ {\d r^2\over 1-2m(r)/r} + r^2\d\Omega^2; 
\end{equation}
then
\begin{equation}
\d \lambda \propto \sqrt{ e^{2\Phi(r)}\{\d r^2 + [1-2m(r)/r] r^2 \d\Omega^2\}}.
\end{equation}
So in particular, for radial null geodesics, the affine parameter is 
\begin{equation}
\d\lambda\propto e^{\Phi(r)} \, \d r; \qquad
\lambda \propto \int e^{\Phi(r)} \, \d r.
\end{equation}
This result does not appear to be particularly well-known (but see, for instance,~\cite{Lorentzian}).

\item Static spherical symmetry (in Buchdahl coordinates):\\
One has~\cite{Buchdahl,Buchdahl2,Buchdahl3}
\begin{equation}
\d s^2 = - f(r) \d t^2 
+ {\d r^2\over f(r)} + \Sigma(r)^2 \,\d\Omega^2; 
\end{equation}
then
\begin{equation}
\d \lambda \propto \sqrt{ \d r^2 + f(r) \,\Sigma(r)^2 \,\d\Omega^2}.
\end{equation}
So in particular, for radial null geodesics in Buchdahl coordinates, the affine parameter is simply the radial coordinate
\begin{equation}
\d\lambda\propto \d r; \qquad
\lambda \propto   r.
\end{equation} \enlargethispage{20pt}
\hspace{-2pt}This result does not appear to be at all well-known. 

\item
Exponential metric: \\
One has~\cite{exponential}
\begin{equation}
\d s ^2 = -e^{-2m/r} \d t^2 + e^{2m/r}\{ \d r^2 + r^2\, \d\Omega^2 \};
\end{equation}
then 
\begin{equation}
\d \lambda \propto \sqrt{ \d r^2 + r^2\d\Omega^2}.
\end{equation}
So for the exponential metric, the null affine parameter is simply the naive distance $\d s_0 =\sqrt{ \d r^2 + r^2\d\Omega^2}$. 
This result does not appear to be particularly well-appreciated. 
\item
Black bounce spacetimes:\\
One has~\cite{SV, Bronnikov:2021}
\begin{equation}
\d s ^2 = -\left(1-{2m\over\sqrt{r^2+a^2}} \right) \d t^2 + { \d r^2 \over 1-{2m\over\sqrt{r^2+a^2}} } + (r^2+a^2)\, \d\Omega^2 \};
\end{equation}
then 
\begin{equation}
\d \lambda \propto \sqrt{ \d r^2 +\left(1-{2m\over\sqrt{r^2+a^2}} \right) (r^2+a^2) \d\Omega^2}.
\end{equation}
So for the black bounce metric, the radial null affine parameter is simply $\lambda \propto r$. \\
This result does not appear to be particularly well-appreciated. 

\item More generally, consider Synge's ``confromastat'' metrics~\cite{Synge} (signifying: static, spatially conformally flat).  
The line element can then without a loss of generality be cast in the form
\begin{equation}
\d s^2 = - e^{2\Phi_0(\vec x)}\, \d t^2 + e^{2\Phi_1(\vec x)}\,\d\vec x \cdot \d\vec x.
\end{equation}
That is, the spacetime metric can be put into the form
\begin{equation}
g_{ab} = \left[ \begin{array}{c|c} - e^{2\Phi_0(\vec x)} & 0 \\ 
\hline
0 & e^{2\Phi_1(\vec x)}\, \delta_{ij}
\end{array} \right].
\end{equation}

\clearpage
Then, the affine three-metric for these conformastat spacetimes is itself conformally flat:
\begin{equation}
h_{ij} =  e^{2[\Phi_0(\vec x)+\Phi_1(\vec x)]} \, \delta_{ij},
\end{equation}
corresponding to
\begin{equation}
\d \lambda =  e^{[\Phi_0(\vec x)+\Phi_1(\vec x)]} \; ||\d \vec x||. 
\end{equation}
So for the conformastat metrics, the null affine parameter is simply proportional (in a position-dependent manner) to the naive distance $\d s_0 = ||\d \vec x||$. 

\item 
A special case of the conformastat metrics is $\Phi_0(\vec x)=-\Phi_1(\vec x)$
(this is a generalization of the exponential metric mentioned above).
The line element can then without a loss of generality be cast in the form
\begin{equation}
\d s^2 = - e^{2\Phi(\vec x)}\, \d t^2 + e^{-2\Phi(\vec x)}\,\d\vec x \cdot \d\vec x.
\end{equation}
That is, the spacetime metric can be put into the form
\begin{equation}
g_{ab} = \left[ \begin{array}{c|c} - e^{2\Phi(\vec x)} & 0 \\ 
\hline
0 & e^{-2\Phi(\vec x)}\, \delta_{ij}
\end{array} \right].
\end{equation}
Then, the affine three-metric for this specific subclass of conformastat spacetimes is flat:
\begin{equation}
h_{ij} = \delta_{ij},
\qquad
\d \lambda =  ||\d \vec x||. 
\end{equation}
So for this specific subclass of conformastat metrics, the null affine parameter is simply the naive distance $\d s_0 = ||\d \vec x||$. 
\end{itemize}


%
Many other manifestly static examples could be considered, but instead let us now consider the mathematically more general, and astrophysically more important, case of stationary (nonstatic) spacetimes.

\section{Stationary spacetimes}

A stationary spacetime is characterized by the global existence of a Killing vector $K$ (not necessarily hypersurface orthogonal), which is timelike (that is, $g(K,K)<0$) in the domain of outer communication.  See references~\cite{Kerr:1963,Kerr:1965,Kerr:2007,Kerr-book, Kerr-intro, 
Kerr-book2, Newman:1965a, Newman:1965b, Doran:1999, Medved:2004, Liberati:2018, Kerr-ansatz, Teukolsky:2014, Adamo:2014}.

\subsection{ADM Form}
In any stationary spacetime, one can without a loss of generality choose coordinates such that the metric is of ADM form,
see references~\cite{MTW,Wald, Carroll, Hartle, Schutz1,Weinberg1,Weinberg2, Peebles, Padmanabhan, Hobson,Sachs,Stephani,Dinverno, Poisson, Lorentzian,ADM,Gourgoulhon:2007,Hawking:1995}, and all the metric components are time-independent
\begin{equation}
g_{ab} = \left[\begin{array}{c|c}
- N^2 + g^{kl} v_k v_l & v_i \\ \hline v_j & g_{ij}
\end{array} \right].
\end{equation}
{For completeness, note 
\begin{equation}
g^{ab} = \left[\begin{array}{c|c}
- N^{-2} & v_i/N^2 \\ \hline v_j/N^2 & g^{ij}- v^i v^j/N^2
\end{array} \right].
\end{equation}
Here, $g^{ij}= [g_{ij}]^{-1}$ and $v^i = g^{ij} v_j$. 
}
Then, for any null curve
\begin{equation}
\label{E:null}
- (N^2-  g^{kl} v_k v_l) \d t^2 + 2 v_i \d x^i \d t+ g_{ij} \d x^i \d x^j = 0.
\end{equation}
For any affinely parameterized null geodesic, one has the conservation law
\begin{equation}
\label{null-stat}
g\left(K, {\d x^a \over \d \lambda } \right) = 
g_{ta} {\d x^a \over \d \lambda } 
= - (N^2-  g^{kl} v_k v_l) {\d t \over \d \lambda } 
+ v_i {\d x^i \over \d \lambda }  = \hbox{constant} = \epsilon.
\end{equation}
Let us rewrite this as
\begin{equation}
-\sqrt{N^2-  g^{kl} v_k v_l} \; {\d t \over \d \lambda } 
+ {v_i\over \sqrt{N^2-  g^{kl} v_k v_l}}\,  {\d x^i \over \d \lambda }  = { \epsilon\over \sqrt{N^2-  g^{kl} v_k v_l}}. 
\end{equation}
Squaring both sides:
\begin{equation}
(N^2-  g^{kl} v_k v_l) \left({\d t \over \d \lambda } \right)^2 
- 2 v_i {\d x^i\over \d \lambda} {\d t \over \d \lambda }
+ {\left(v_i {\d x^i\over \d\lambda}\right)^2\over (N^2-  g^{kl} v_k v_l)}
 = {\epsilon^2 \over (N^2-  g^{kl} v_k v_l)}.
\end{equation}
That is
\begin{equation}
\label{E:cons}
(N^2-  g^{kl} v_k v_l) {\d t }^2 
- 2 v_i {\d x^i} {\d t  }
+ {\left(v_i {\d x^i}\right)^2\over (N^2-  g^{kl} v_k v_l)}
 = {\epsilon^2 \over (N^2-  g^{kl} v_k v_l)} \d \lambda^2.
\end{equation}
Now add the two Equations (\ref{E:null}) and (\ref{E:cons}) derived above
to obtain
\begin{equation}
 g_{ij} \d x^i \d x^j + { \left(v_i {\d x^i}\right)^2 \over (N^2-  g^{kl} v_k v_l)}
= {\epsilon^2 \over (N^2-  g^{kl} v_k v_l)} \d \lambda^2.
\end{equation}
Rearranging
\begin{equation}
\epsilon^2\d\lambda^2 = 
(N^2-  g^{kl} v_k v_l) g_{ij} \d x^i \d x^j +  \left(v_i {\d x^i}\right)^2.
\end{equation}
That is
\begin{equation}
\epsilon^2\d\lambda^2 = 
\{ (N^2-  g^{kl} v_k v_l) g_{ij} + v_i v_j \} \; \d x^i \d x^j. 
\end{equation}
Thence, along any affinely parameterized null geodesic in any stationary spacetime, one has
\begin{equation}
\d \lambda \propto 
\sqrt{ \left\{ (N^2-  g^{kl} v_k v_l) g_{ij} + v_i v_j \right\}  \; \d x^i \d x^j}.
\end{equation}
Defining the affine three-metric
\begin{equation}
h_{ij} = (N^2-  g^{kl} v_k v_l) g_{ij} + v_i v_j= - g_{tt} g_{ij} + g_{ti} g_{tj},
\end{equation}
we see that for any affinely parameterized null geodesic in any stationary spacetime, one has
\begin{equation}
\d \lambda \propto 
\sqrt{ h_{ij}   \, \d x^i \,\d x^j}.
\end{equation}
{We emphasize: the affine three-metric is very simply determined by the lapse, shift, and physical three-metric appearing in the ADM decomposition.
}

\subsection{Threaded Form}
Alternatively, one can choose to put the metric in {so-called} ``threaded'' (Kaluza--Klein-like) form~\cite{thread,vanElst:1996,Ellis:1998}:
\begin{equation}
\d s^2 = - V^2 (\d t + \omega_i \,\d x^i)^2 + \gamma_{ij} \d x^i \d x^j.
\end{equation}
That is,
\begin{equation}
g_{ab} = \left[ \begin{array}{c|c} - V^2 & - V^2 \omega_j \\ 
\hline
-V^2 \omega_i & \gamma_{ij} - V^2 \omega_i \omega_j 
\end{array} \right].
\end{equation}
{Qualitatively similar metrics, though {typically} in higher space--time dimensions, commonly arise in Kaluza--Klein theory~\cite{Kaluza:1921, Klein:1926, Overduin:1997, Witten:1981, Salam:1981}.}
{For completeness, note 
\begin{equation}
g^{ab} = \left[ \begin{array}{c|c} - V^{-2}+ g^{ij} \omega_i\omega_j & -  \omega^j \\ 
\hline
- \omega^i & \gamma^{ij} 
\end{array} \right].
\end{equation}
Here, $\gamma^{ij}= [\gamma_{ij}]^{-1}$ and $\omega^i = \gamma^{ij} \omega_j$. 
That is, the threaded form of the metric is in some sense dual to the ADM form of the metric.
}

Then, in threaded form, the affine three-metric is particularly simple:
\begin{equation}
h_{ij} = - g_{tt} g_{ij} + g_{ti} g_{tj} = V^2 (\gamma_{ij} - V^2 \omega_i \omega_j ) 
+ (V^2 \omega_i) (V^2 \omega_j) = V^2 \gamma_{ij}. 
\end{equation}
{
Thence,
\begin{equation}
\d \lambda \propto 
|V| \sqrt{ \gamma_{ij}   \, \d x^i \,\d x^j}.
\end{equation}
{We emphasize: the affine three-metric is very simply determined by the quantities appearing in the threading decomposition.}

These two analyses completely settle the stationary case.  
}

\subsection{{Examples} 
}
\begin{itemize}
\item Strong Painlev\'e--Gullstrand metrics (static but not manifestly so):\\
{In this situation, one} has~\cite{Martel:2000,Volovik:1999, PG,PG-paddy} 
{
\begin{equation}
g_{ab} = \left[\begin{array}{c|c}
- 1 + \delta^{kl} v_k v_l & v_i \\ \hline v_j & \delta_{ij}
\end{array} \right].
\end{equation}
Thence,}
\begin{equation}
g_{ij} \to \delta_{ij}; \qquad N\to 1; \qquad
\d \lambda \propto 
\sqrt{ \{ (1-  \delta^{kl} v_k v_l) \delta_{ij} + v_i v_j \}  \; \d x^i\, \d x^j}.
\end{equation}
That is,
\begin{equation}
\d \lambda \propto 
\sqrt{  \left( \delta_{ij} -  |v|^2 \{\delta_{ij} - \hat v_i \hat v_j \}  \right)\; \d x^i\, \d x^j}.
\end{equation}
Then, for null geodesics parallel to the flow $\hat v$, we have $\d\lambda \propto 
\sqrt{  \delta_{ij} \,\d x^i \,\d x^j}$, while for null geodesics perpendicular to the flow $\hat v$, we have $\d\lambda \propto 
\sqrt{ -g_{tt}  \,\delta_{ij} \,\d x^i \,\d x^j}$.
{That is,  $\d\lambda \propto \sqrt{1-|v|^2} \sqrt{\,\delta_{ij} \,\d x^i \,\d x^j}$.}

\item Weak Painlev\'e--Gullstrand metrics (static but not manifestly so):\\
{In this situation, one} has~\cite{Martel:2000,Volovik:1999, PG,PG-paddy} 
{
\begin{equation}
g_{ab} = \left[\begin{array}{c|c}
- N^2 + \delta^{kl} v_k v_l & v_i \\ \hline v_j & \delta_{ij}
\end{array} \right].
\end{equation}
Thence,}
\begin{equation}
g_{ij} \to \delta_{ij}; \qquad 
\d \lambda \propto 
\sqrt{ \{ (N^2-  \delta^{kl} v_k v_l) \delta_{ij} + v_i v_j \}  \; \d x^i \d x^j}.
\end{equation}
That is
\begin{equation}
\d \lambda \propto 
\sqrt{  \left( N^2 \delta_{ij} -  |v|^2 \{\delta_{ij} - \hat v_i \hat v_j \}  \right)\; \d x^i \d x^j}.
\end{equation}
Then, for null geodesics parallel to the flow $\hat v$, we have $\d\lambda \propto 
N \sqrt{  \delta_{ij} \,\d x^i \,\d x^j}$, while for null geodesics perpendicular to the flow $\hat v$, we have $\d\lambda \propto 
\sqrt{ -g_{tt}  \,\delta_{ij} \,\d x^i \,\d x^j}$.
{That is,  $\d\lambda \propto \sqrt{N^2-|v|^2} \sqrt{\,\delta_{ij} \,\d x^i \,\d x^j}$.}

\item Boyer--Lindquist metrics (not necessarily limited to Kerr or Kerr--Newman {spacetime}):\\
One has, see references~\cite{Kerr:1963,Kerr:1965,Kerr:2007,Kerr-book, Kerr-intro, Kerr-book2, Newman:1965a, Newman:1965b, Doran:1999, Medved:2004, Liberati:2018, Kerr-ansatz, Teukolsky:2014, Adamo:2014,Boyer:1966, Baines:2020, Rajan:2016, Schuster:2018, Bambi:2013},
\begin{equation}
g_{ij} = \left[\begin{array}{ccc} g_{rr}&0&0\\ 0& g_{\theta\theta} & 0 \\
0 &0&g_{\phi\phi} \end{array} \right]_{ij};   \qquad v_i = (0,0,v_\phi)= (0,0,g_{t\phi})
\end{equation}
Then,
\begin{equation}
\d \lambda \propto 
\sqrt{ \left\{ - g_{tt} \;  g_{ij} + v_i v_j \right\}  \; \d x^i \, \d x^j}.
\end{equation}
That is,
\begin{equation}
\d \lambda \propto 
\sqrt{ \left[
\begin{array}{ccc}  
- g_{tt} g_{rr}&0&0\\ 
0&  - g_{tt} g_{\theta\theta} & 0 \\
0 &0& - g_{tt}  g_{\phi\phi} + g_{t\phi}^2
\end{array} 
\right]_{ij} \; \d x^i \,\d x^j}.
\end{equation}
The affine three-metric is  singular at both the ergosurfaces, where $g_{tt}=0$, and at the horizons, where $ g_{tt}  g_{\phi\phi} - g_{t\phi}^2=0$. 

\item For the Kerr spacetime itself, the affine three-metric is particularly simple. A brief calculation (in Boyer--Lindquist coordinates) yields
\begin{equation}
\d \lambda \propto 
\sqrt{ \left[
\begin{array}{ccc}  
{r^2-2mr+a^2\cos^2\theta\over r^2-2mr+a^2}&0&0\\ 
0&  \scriptstyle{r^2-2mr+a^2\cos^2\theta}& 0 \\
0 &0& \scriptstyle{(r^2-2mr+a^2)\sin^2\theta}
\end{array} 
\right]_{ij} \; \d x^i \, \d x^j}.
\end{equation}
{
This can also be written as
\begin{equation}
\d \lambda \propto 
\sqrt{ \left[
\begin{array}{ccc}  
1-{a^2\sin^2\theta\over r^2-2mr+a^2}&0&0\\ 
0&  \scriptstyle{r^2-2mr+a^2\cos^2\theta}& 0 \\
0 &0& \scriptstyle{(r^2-2mr+a^2)\sin^2\theta}
\end{array} 
\right]_{ij} \; \d x^i \, \d x^j}.
\end{equation}
}
\hspace{-2pt}It is somewhat unusual to have any Kerr-related result looking relatively simple.\\
In particular, for photons emitted along the axis of rotation, we have $\lambda \propto r$, as we saw happens for Schwarzschild.
{More subtly, for null geodesics confined to the equatorial plane $\theta=\pi/2$, we can restrict the spatial coordinates to $x^i = (t,\phi)$ and introduce an affine two-metric
\begin{equation}
\d \lambda \propto 
\sqrt{ \left[
\begin{array}{cc}  
1-{a^2\over r^2-2mr+a^2}&0\\ 
0& \scriptstyle{(r^2-2mr+a^2)}
\end{array} 
\right]_{ij} \; \d x^i \, \d x^j}.
\end{equation}

}

\end{itemize}

Many other examples of stationary spacetimes could be considered, but let us now turn to some time-dependent situations.


\clearpage
\section{Spacetimes with timelike conformal Killing vector}

Suppose now that we have a preferred timelike vector field $K^a$, which we do not necessarily insist is a Killing vector. 

\subsection{ADM Form}
Consider the timelike integral curves of $K$ and take a spatial slice transverse to them; note that we only require transversality, not orthogonality. Then, without any further loss of generality, let us set up an ADM coordinate system in which both
\begin{equation}
g_{ab}(t,\vec x) = \left[\begin{array}{c|c}
- N^2 + g^{kl} v_k v_l & v_i \\ \hline v_j & g_{ij}
\end{array} \right],
\end{equation}
and
\begin{equation}
K^a(t,\vec x) = ( e^\Psi;0,0,0).
\end{equation}
(Here, all of these quantities can now be both space- and time-dependent).

Let us now assume $K^a$ is a conformal Killing vector, $K_{(a;b)} = \L_K g_{ab} \propto g_{ab}$. 
We will not need to know what the proportionality factor is.
That is,
\begin{equation}
K^c \partial_c g_{ab} + (\partial_a K^c) g_{cb} +(\partial_b K^c) g_{ac}\propto g_{ab}.
\end{equation}
By constructing $\partial_a K^c = (\partial_a \Psi) K^c$, the conformal Killing equation reduces to
\begin{equation}
K^c \partial_c g_{ab} + (\partial_a \Psi) K_{b} +
K_a (\partial_b \Psi) \propto g_{ab}.
\end{equation}
Now for convenience, define $T^a=(1;0,0,0)$ so that $K^a = e^\Psi \; T^a$; we have
\begin{equation}
\partial_t g_{ab} + (\partial_a \Psi) T_{b} +
T_a (\partial_b \Psi) g_{ac}\propto g_{ab}.
\end{equation}
As long as $K^a$ is a conformal Killing vector, $K_{(a;b)} = \L_K g_{ab} \propto g_{ab}$, then we still have a conservation law
for any affinely parameterized null geodesic:
\begin{equation}
K^a g_{ab} {\d x^b \over \d \lambda } =
 e^\Psi g_{tb} {\d x^b \over \d \lambda } 
= e^\Psi\left\{ - (N^2-  g^{kl} v_k v_l) {\d t \over \d \lambda } 
+ v_i {\d x^i \over \d \lambda } \right\}  = \hbox{constant} = \epsilon.
\end{equation}
{This is a slight modification to Equation (\ref{null-stat}) above, now allowing for a nontrivial conformal distortion $e^\Psi$.}
Thence,
\begin{equation}
\left\{ - (N^2-  g^{kl} v_k v_l) {\d t \over \d \lambda } 
+ v_i {\d x^i \over \d \lambda } \right\}  = e^{-\Psi} \epsilon.
\end{equation}
Now, carry through essentially the same algebra as for the stationary case.

We still have
\begin{equation}
- (N^2-  g^{kl} v_k v_l) \d t^2 + 2 v_i \d x^i \d t+ g_{ij} \d x^i \d x^j = 0,
\end{equation}
but now 
\begin{equation}
(N^2-  g^{kl} v_k v_l) \left({\d t } \right)^2 
- 2 v_i {\d x^i} {\d t}
+{ \left(v_i {\d x^i}\right)^2 \over (N^2-  g^{kl} v_k v_l)}
= {\epsilon^2 e^{-2\Psi}\over (N^2-  g^{kl} v_k v_l)} \d \lambda^2.
\end{equation}
This implies
\begin{equation}
 g_{ij} \d x^i \d x^j + { \left(v_i {\d x^i}\right)^2 \over (N^2-  g^{kl} v_k v_l)}
= {\epsilon^2 e^{-2\Psi} \over (N^2-  g^{kl} v_k v_l)} \d \lambda^2,
\end{equation}
which we can rewrite as
\begin{equation}
\epsilon^2\d\lambda^2 = 
e^{2\Psi}\left\{ (N^2-  g^{kl} v_k v_l) g_{ij} \d x^i \d x^j +  \left(v_i {\d x^i}\right)^2\right\}.
\end{equation}
That is,
\begin{equation}
\epsilon^2\d\lambda^2 = 
e^{2\Psi}\{ (N^2-  g^{kl} v_k v_l) g_{ij} + v_i v_j \} \; \d x^i \d x^j.
\end{equation}
Therefore, along any affinely parameterized null geodesic, one still has
\begin{equation}
\d \lambda \propto 
\sqrt{ e^{2\Psi}\{ (N^2-  g^{kl} v_k v_l) g_{ij} + v_i v_j \}  \; \d x^i \,\d x^j}
\end{equation}
This can also be written as
\begin{equation}
\d \lambda \propto 
\sqrt{ e^{2\Psi}\{  - g_{tt} \,g_{ij} + g_{ti}\, g_{tj} \}  \; \d x^i \,\d x^j}
\end{equation}
The affine three-metric is now
\begin{equation}
h_{ij} =  e^{2\Psi}\{  - g_{tt} \, g_{ij} + g_{ti} \, g_{tj} \}.
\end{equation}
Note the extra conformal factor $e^{2\Psi}$ now appearing in the affine three-metric.
{Note that this conformal factor $e^{2\Psi}$ is not coming directly from the spacetime metric; it is coming indirectly from the assumed conformal Killing vector.}

\subsection{Threaded Form}
Alternatively, one can choose to put the metric in ``threaded''~\cite{thread, vanElst:1996, Ellis:1998} (Kaluza--Klein-like) 
form:
\begin{equation}
g_{ab} = \left[ \begin{array}{c|c} - V^2 & - V^2 \omega_j \\ 
\hline
-V^2 \omega_i & \gamma_{ij} - V^2 \omega_i \omega_j 
\end{array} \right].
\end{equation}
Then, in threaded form,  the affine three-metric is particularly simple:
\begin{equation}
h_{ij} = e^{2\Psi} \{- g_{tt} g_{ij} + g_{ti} g_{tj} \} = 
e^{2\Psi} \{V^2 (\gamma_{ij} - V^2 \omega_i \omega_j ) 
+ (V^2 \omega_i) (V^2 \omega_j) \} = e^{2\Psi} V^2 \gamma_{ij}. 
\end{equation}
That is, 
\begin{equation}
h_{ij} = e^{2\Psi} V^2 \gamma_{ij}. 
\end{equation}
Again, note the extra conformal factor $e^{2\Psi}$  now appearing in the affine three-metric.
{Again, note that this conformal factor $e^{2\Psi}$ is not coming directly from the spacetime metric; it is coming indirectly from the assumed conformal Killing vector.}

\subsection{{Examples }} 
\enlargethispage{15pt}
\begin{itemize}
\item 
FLRW cosmology:
\begin{equation}
\d s^2 = - \d t^2 + a(t)^2 \left\{ {\d r^2\over 1- kr^2} + r^2 \d\Omega^2 \right\}; \qquad K^a = (a(t);0,0,0).
\end{equation}
Then
\begin{equation}
\d \lambda \propto 
\sqrt{ a(t)^4 \left[ {\d r^2\over 1- kr^2} + r^2 \d\Omega^2\right] \ }  = a(t)^2  \; \d s_0
= a(t) \d t,
\end{equation}
where $s_0$ is the distance in the normalized hypersphere/flat/hyperbolic plane with a spatial three-metric $[g_0]_{ij}$.
The affine three-metric can be written as
\begin{equation}
h_{ij}(t,\vec x) = a(t)^4 \; [g_0]_{ij}(x). 
\end{equation}

To check this, work in conformal time $\d t = a(\eta) \d\eta$ wherein
\begin{equation}
\d s^2 =  a(\eta)^2 \left\{-\d\eta^2 + {\d r^2\over 1- kr^2} + r^2 \d\Omega^2 \right\}; \qquad K^a = (1;0,0,0);
\end{equation}
{(working in these coordinates, it is particularly easy to check that $K$ is indeed a conformal Killing vector).}
Then,
\begin{equation}
\d \lambda \propto 
\sqrt{ a(\eta)^4 \left[ {\d r^2\over 1- kr^2} + r^2 \d\Omega^2\right] \ }
= a(\eta)^2 \d \eta = a(t) \d t.
\end{equation}
\enlargethispage{15pt}

\item Any constant-spatial-shape cosmology:
\begin{equation}
\d s^2 = - \d t^2 + a(t)^2 \left\{ [g_0]_{ij}(\vec x) \,\d x^i \,\d x^j\right\}; 
\qquad K^a = (a(t);0,0,0).
\end{equation}
{(the key constraint here is that the shape of the spatial slices is assumed to be governed by  $[g_0]_{ij}(\vec x)$, which  is assumed to be an arbitrary time-independent three-metric).}
Then,
\begin{equation}
\d \lambda \propto 
\sqrt{ a(t)^4 \left\{[g_0]_{ij}(\vec x) \,\d x^i\,\d x^j\right\}  }= a(t)^2 \, \d s_0 = a(t) \d t,
\end{equation}
where $s_0$ is the distance in the normalized constant-spatial-shape slices with the spatial three-metric $[g_0]_{ij}$ .

The affine three-metric can be written as
\begin{equation}
h_{ij}(t,\vec x) = a(t)^4 \; [g_0]_{ij}(x). 
\end{equation}

To check this work in conformal time,  $\d t = a(\eta) \d\eta$, wherein
\begin{equation}
\d s^2 =  a(\eta)^2 \left\{-\d\eta^2 +[g_0]_{ij}(\vec x) \,\d x^i\, \d x^j \right\}; \qquad K^a = (1;0,0,0);
\end{equation}
{(working in these coordinates, it is particularly easy to check that $K$ is indeed a conformal Killing vector).}
Then,
\begin{equation}
\d \lambda \propto 
\sqrt{ a(\eta)^4 \left\{ [g_0]_{ij}(\vec x) \,\d x^i \,\d x^j]\right\} }
= a(\eta)^2 \,\d\eta = a(t) \,\d t.
\end{equation}
\end{itemize}
Many other examples along these lines could be developed.

\section{Geometrical (3 + 1) Interpretation}\enlargethispage{20pt}

Can we give the affine three-metric $h_{ij}$ a geometrical (3 + 1) spacetime interpretation?
Consider the (3 + 1) spacetime metric
\begin{equation}
g_{ab}(x,t) = \left[\begin{array}{c|c}
- N^2 + g^{kl} v_k v_l & v_i \\ \hline v_j & g_{ij}
\end{array} \right],
\end{equation}
with
\begin{equation}
K^a(x,t) = ( e^\Psi;0,0,0),
\end{equation}
and define the four-index tensor
\begin{equation}
Z_{abcd} = g_{ac} \,g_{bd} - g_{ad} \, g_{bc}.
\end{equation}
This tensor shows up quite naturally in curved spacetime electromagnetism.

Now  define the two-index tensor $h_{ab}$ obtained by projecting the four-index tensor $Z_{abcd}$ onto the spatial hyperplanes orthogonal to the conformal timelike Killing vector: 
\begin{equation}
h_{ac} = Z_{abcd}\, K^b\, K^d = g_{ac}(g_{bd} K^b K^d) - (g_{ad} K^d) (g_{bc} K^c).
\end{equation}
Here, 
\begin{equation}
(g_{bd} K^b K^d)  = e^{2\Psi} g_{tt} ; 
\qquad \hbox{and} \qquad
(g_{ad} K^d) = e^\Psi (g_{tt}, v_i).
\end{equation}
Thence,
\begin{eqnarray}
h_{cd}&=& Z_{abcd} K^b K^d 
\nonumber\\
&=& e^{2\Psi} \left\{ g_{tt} g_{ac} - (g_{tt}, v_i)_a (g_{tt}, v_j)_c \right\}
\nonumber\\
&=& e^{2\Psi} \left[\begin{array}{c|c}
0& 0 \\ \hline 0 & g_{tt} g_{ij} - v_i v_j
\end{array} \right]_{ac} 
\nonumber\\
&=&
 -  \left[\begin{array}{c|c}
0& 0 \\ \hline 0 & h_{ij}
\end{array} \right]_{ac}.
\end{eqnarray}
This gives a (3 + 1) spacetime perspective on just where the affine three-metric 
\begin{equation}
h_{ij} \propto e^{2\Psi} \{-g_{tt} g_{ij} + v_i v_j\} = e^{2\Psi} \{-g_{tt} g_{ij} + g_{ti} g_{tj}\}
= e^{2\Psi}  V^2 \gamma_{ij}
\end{equation}
was coming from.\enlargethispage{20pt}
Note that
\begin{equation}
\L_K h_{ac} = (\L_K Z_{abcd} ) K^b K^c \propto Z_{abcd}  K^b K^c = h_{ac},
\end{equation}
so $K^b$ is also a conformal Killing vector of the singular metric $h_{ac}$.

\section{Conformal transformations}

Now consider an arbitrary spacetime with metric $ \hat g_{ab}$ and no particular symmetries. 
Under the conformal transformation $g_{ab} = e^{2\Theta} \, \hat g_{ab}$, we have
\begin{equation}
\Gamma^a{}_{bc} = \hat \Gamma^a{}_{bc} 
+ \left(\delta^a{}_b \Theta_{,c} + \delta^a{}_c \Theta_{,b} - \Theta^{,a} \hat g_{bc}\right).
\end{equation}
Thence, if $\hat k$ is an affinely parameterized null geodesic, $\hat\nabla_{\hat k} \hat k=0$, we have
\begin{equation}
[\nabla_{\hat k} \hat k]^a 
= \left(\delta^a{}_b \,\Theta_{,c} + \delta^a{}_c \,\Theta_{,b} 
- \Theta^{,a} \,\hat g_{bc}\right) \hat k^b \hat k^c
= 2 (\hat k^c \Theta_{,c} ) \hat k^a.
\end{equation}
Now define $k$ by $\hat k = f k$, then
\begin{equation}
[\nabla_{f k} (f k)]^a = f^2 [\nabla_{k} k]^a + f (k\cdot \nabla f) k^a = 
2 f^2 (k^c \Theta_{,c} ) k^a
\end{equation}\clearpage
\noindent That is
\begin{equation}
[\nabla_{k} k]^a  = 
2   (k^c \Theta_{,c} ) k^a-    (f^{-1} k\cdot \nabla f )k^a
\end{equation}
Thence, $k$ will be affinely parameterized with respect to $g$ provided we pick $f=e^{2\Theta}$. This is equivalent to setting $d\lambda/\d\hat\lambda = e^{2\Theta}$, or more explicitly
\begin{equation}
\d\lambda = e^{2\Theta} \d\hat\lambda.
\end{equation}
Even more explicitly: \emph{if} we can construct an ``affine 3-metric'' $\hat h_{ij}$ for the original spacetime metric $\hat g_{ab}$ so that
\begin{equation}
\d \hat \lambda = \sqrt{ \hat h_{ij}\;\d x^i \; \d x^j}
\end{equation}
\emph{then} for the conformally transformed spacetime metric 
$g_{ab} = e^{2\Theta} \; \hat g_{ab}$, we have \footnote{Note the perhaps naively unexpected occurrence of the square $ e^{4\Theta}$  of the conformal factor $e^{2\Theta}$. }
\begin{equation}
 h_{ij} = e^{4\Theta}\;\hat h_{ij}.
\end{equation}
In particular, given our previous results, this construction will work for any spacetime that is conformal to a spacetime containing a timelike conformal Killing vector.


\subsection{Examples}

\begin{itemize}
\item
As a consistency check, note that the observations above are  compatible with what we already saw happening for FLRW spacetimes and for constant-spatial-shape cosmologies.

\item CFLRW spacetimes: As a further generalization, consider the conformally FLRW spacetimes of reference~\cite{CFLRW}:
\begin{equation}
\d s^2 =  e^{2\Theta(x)} \left[ - \d t^2 + a(t)^2 \left\{ {\d r^2\over 1- kr^2} + r^2 \d\Omega^2 \right\}\right].
\end{equation}
{Here, $\Theta(x)$ is allowed to depend on both space and time.} 
The CFLRW spacetimes are of interest because they have the same null geodesics (though not the same null affine parameter) as FLRW. 
In view of the preceding discussion, for any such spacetime, we can identify a suitable affine parameter as
\begin{equation}
\d \lambda = \sqrt{ \hat h_{ij}\;\d x^i \; \d x^j} \propto 
\sqrt{  e^{4\Theta(x)}  \; a(t)^4\;  \left\{ {\d r^2\over 1- kr^2} + r^2 \d\Omega^2 \right\}}.
\end{equation}
We can also unwrap this as 
\begin{equation}
\d \lambda  \propto  e^{2\Theta(x)}  \, a(t) \, \d t.
\end{equation}
\end{itemize}

More examples along these lines could be developed.

\section{General Case}
\def\Z{{\mathcal{Z}}}
The completely general case is considerably more subtle, and we can at best develop an implicit nonconstructive formalism. As long as there was at least some symmetry---at least having the  spacetime conformally related to a spacetime containing a conformal timelike Killing vector---then we have already demonstrated the existence of an affine three-metric $h_{ij}$  completely characterizing the possible null affine parameters. If we try to dispense with symmetries altogether, then the best that can be conducted is an implicit Finsler-like construction~\cite{Bekenstein:1992, Perlick:2005, Gibbons:2007, Gibbons:2008, Skakala:2010, Lammerzahl:2012, Lammerzahl:2018, Pfeifer:2019}. 
Indeed, if we write 
\begin{equation}
g_{ab}(x,t) = \left[\begin{array}{c|c}
- N^2 + g^{kl} v_k v_l & v_i \\ \hline v_j & g_{ij}
\end{array} \right],
\end{equation}
then for any null curve, we still have
\begin{equation}
\label{E:null2}
- (N^2-  g^{kl} v_k v_l) \d t^2 + 2 v_i \,\d x^i \,\d t+ g_{ij} \,\d x^i \,\d x^j = 0.
\end{equation}
Can we now argue for some generalization of our previous construction? Perhaps something along the lines of 
\begin{equation}
\d \lambda \propto 
\sqrt{ \hbox{(something)} \; \{  - g_{tt} g_{ij} + v_i v_j \}  \; \d x^i \,\d x^j}\; 
= \sqrt{ \hbox{(something)} \; \{  V^2 \gamma_{ij} \}  \; \d x^i \,\d x^j}\; 
?
\end{equation}
Indeed, let $T^a=(1,0,0,0)$, and
for any affinely parameterized null geodesic, define
\begin{equation}
\Z(x, \d x/\d\lambda) = T^a g_{ab} {\d x^b \over \d \lambda } =
 g_{tb} {\d x^b \over \d \lambda } 
=\left\{ - (N^2-  g^{kl} v_k v_l) {\d t \over \d \lambda } 
+ v_i {\d x^i \over \d \lambda }\right\}.
\end{equation}
We do not know how to calculate $\Z(x, \d x/\d\lambda)$ explicitly, but we know it must exist. Unfortunately, in the general case (without appealing to any Killing symmetries), this quantity $\Z(x, \d x/\d\lambda)$ can depend on the specific null geodesic in question; it is no longer a function on the manifold, but is instead a function on the tangent bundle.

Then, similarly to what we saw for the (conformal) timelike Killing vectors, we see
\begin{equation}
{\Z(x, \d x/\d\lambda) \d\lambda\over\sqrt{(N^2-  g^{kl} v_k v_l)}}
=
\left\{ - \sqrt{N^2-  g^{kl} v_k v_l} \d t
+ {v_i\over\sqrt{N^2-  g^{kl} v_k v_l}}  \d x^i \right\}.
\end{equation}
Squaring both sides, this then implies
\begin{equation}
\left({\Z(x, \d x/\d\lambda) \d\lambda\over\sqrt{(N^2-  g^{kl} v_k v_l)}}\right)^2 
=
(N^2-  g^{kl} v_k v_l) \d t^2 - 2 v_i \,\d x^i \, \d t
+ {v_i v_j \d x^i \d x^j\over N^2-  g^{kl} v_k v_l} 
\end{equation}
Add this to the null cone condition (\ref{E:null2}), and we find
\begin{equation}
\left({\Z(x, \d x/\d\lambda) \d\lambda\over\sqrt{(N^2-  g^{kl} v_k v_l)}}\right)^2 
=
+g_{ij} \d x^i \d x^j
+ {v_i v_j \d x^i \d x^j\over N^2-  g^{kl} v_k v_l} 
\end{equation}
Rearranging
\begin{equation}
\Z^2(x, \d x/\d\lambda) \d\lambda^2 = \left\{(N^2-  g^{kl} v_k v_l) g_{ij} + v_i v_j \right\} \d x^i \d x^j
\end{equation}
Then
\begin{equation}
\d \lambda^2 = \Z(x, \d x/\d\lambda)^{-2} \left\{- g_{tt} g_{ij} + g_{ti} g_{tj} \right\} \d x^i \d x^j
\end{equation}
So we (nonconstructively) deduce the \emph{existence} of a Finslerian affine three-metric $h_{ij}$, defined up to a (in general unknown) conformal factor defined over the tangent bundle:
\begin{equation}
h_{ij} = { \{- g_{tt} g_{ij} + g_{ti} g_{tj} \}\over \Z(x, \d x/\d\lambda)^2} 
= {V^2 \gamma_{ij} \over \Z(x, \d x/\d\lambda)^2} . 
\end{equation}
Note that the Finslerian behavior has at least been isolated in the conformal factor---this is the type of structure explored by Bekenstein in reference~\cite{Bekenstein:1992}: a Riemannian metric {(more precisely, a Lorentzian metric)} distorted by a Finslerian conformal factor. 

\enlargethispage{20pt}
Unfortunately, proving the \emph{existence} of this  Finslerian affine three-metric seems to be as far as one can go---without imposing at least some symmetry on the spacetime, it seems to be impossible to explicitly determine (or at the very least, significantly constrain) the Finslerian conformal factor
\begin{equation}
\Z(x, \d x/\d\lambda)=T^a \,g_{ab} \,{\d x^b \over \d \lambda }.
\end{equation}


\enlargethispage{15pt}
\section{Discussion}

In this article, we set up a quite general formalism for the {a priori} identification of affine null parameters in spacetimes conformally related to spacetimes with (at least) a timelike conformal Killing vector. 
The formalism is sufficiently general so as to immediately yield useful information in many physically interesting situations. 
This complements the more traditional {a posteriori} approach, where one first on a case-by-case basis identifies null geodesics and then on a case-by-case basis attempts to find an affine null parameter. A number of explicit examples were provided to illustrate the utility of the procedure.

\bigskip
\bigskip
\hrule\hrule\hrule
\section*{Acknowledgements}

MV was directly supported by the Marsden Fund, \emph{via} a grant administered by the
Royal Society of New Zealand.

\bigskip
\hrule\hrule\hrule
\bigskip

\section*{Abbreviations}

The following abbreviations are used in this manuscript:\\

\noindent 
\begin{tabular}{@{}ll}
ANEC & Averaged null energy condition\\
ADM & Arnowitt--Deser--Misner\\
FLRW & Friedmann--Lemaitre--Robertson--Walker\\
CFLRW & Conformally Friedmann--Lemaitre--Robertson--Walker
\end{tabular}

\bigskip
\hrule\hrule\hrule
\bigskip



\end{document}